\documentclass[preprint,12pt,showkeys,showpacs,nofootinbib,floatfix]{revtex4}
\usepackage{amssymb}
\usepackage{amsfonts}
\usepackage{amsmath}
\usepackage{graphicx}
\usepackage{subfigure}
\usepackage[dvipdfm,
            pdfstartview=FitH,
            bookmarksnumbered=true,
            bookmarksopen=true,
            colorlinks,
            pdfborder=001,
            linkcolor=green,
            anchorcolor=green,
            citecolor=red
            ]{hyperref}
\usepackage[toc,page,title,titletoc,header]{appendix}
\begin{document}
\title{Lifshitz effects on vector condensate induced by a magnetic field}
\author{Ya-Bo Wu$^{1}$}
\thanks{E-mail address:ybwu61@163.com}
\author{Jun-Wang Lu$^{1}$}
\author{Mo-Lin Liu$^{2}$}
\author{Jian-Bo Lu$^{1}$}
\author{Cheng-Yuan Zhang$^{1}$}
\author{Zhuo-Qun Yang$^{1}$}
\affiliation{
$^{1}$Department of Physics, Liaoning Normal University, Dalian, 116029, People's Republic of  China\\
$^{2}$College of Physics and Electronic Engineering, Xinyang Normal University, Xinyang, 464000, People's Republic of China}
\begin{abstract}
 By numerical and analytical methods, we study in detail the effects of the Lifshitz dynamical exponent $z$ on the vector condensate induced by an applied magnetic field in the probe limit.  Concretely, in the presence of the magnetic field,  we obtain the Landau level independent of $z$, and we also find the critical value by coupling a Maxwell complex vector field and an SU(2) field into a (3+1)-dimensional Lifshitz black hole, respectively. The research results show that for the two models with the lowest Landau level, the increasing $z$ improves the response of the critical temperature to the applied magnetic field even without the charge density, and the analytical results uphold the numerical results. In addition, we find that, even in the Lifshitz black hole, the Maxwell complex vector model is still a generalization of  the SU(2) Yang-Mills model. Furthermore, we construct the square vortex lattice and discuss the implications of these results.
\end{abstract}


\maketitle
\section{Introduction}
The gauge/gravity duality~\cite{Maldacena1998,Gubser105,witten253} allows us  to deal with a strongly coupled conformal field theory by using its dual weak gravity. Recently, this holographic correspondence has been widely applied to study the high temperature superconductors that are supposed to involve  the strong interaction.

The holographic $s$-wave superconductor model was first realized via an Einstein-Maxwell complex scalar field in the four-dimensional  Schwarzschild anti-de Sitter(AdS) black hole~\cite{Hartnoll2008}. Using the SU(2) gauge field in the AdS black hole, Ref.~\cite{Gubser2008a} constructed the holographic $p$-wave superconductor. In the $p$-wave model, as the temperature is below a critical value, the black hole becomes instable to developing a vector ``hair" which spontaneously breaks the U(1) symmetry  as well as the spatial rotation symmetry. The vector condensate is dual to a non-trivial vacuum expectation value of the vector operator in the boundary field theory, which therefore mimics the $p$-wave superconductor in the condensed matter system. The holographic $d$-wave model was built up by using a charged spin-two field propagating in the bulk~\cite{Chen:2010mk}. Near the critical point, the holographic superconductor model was also studied via the analytical  Sturm-Liouville (SL) eigenvalue method in Ref.~\cite{Siopsis}.

Recently, a new $p$-wave superconductor model was constructed by coupling a Maxwell complex vector (MCV) field to the four-dimensional Schwarzschild AdS black hole in the probe limit~\cite{Cai:2013pda}. In this model, an applied magnetic field can induce the condensate even without the charge density.  Interestingly, the response of this system to the magnetic field is opposite from the one of ordinary superconductors~\cite{Albash:2008eh,Nakano:2008xc}, but this is quite similar to the case of the QCD vacuum phase transition~\cite{Chernodub:2010qx,Chernodub:2011mc}. Moreover, the triangular vortex lattice structure was reproduced in the $x-y$ plane perpendicular to the magnetic field, which was first observed in Ref.~\cite{Maeda:2009vf}. Taking into account the backreaction of the matter field, this $p$-wave model exhibits a rich phase structure~\cite{Cai:2013aca}, which is then observed from the perspective of the holographic entanglement entropy~\cite{Li:2013rhw}, whereafter, the holographic $p$-wave insulator/superconductor phase transition was modeled by introducing such a MCV field into the five-dimensional AdS soliton in the probe limit~\cite{Cai:2013kaa}. It was shown that the Einstein MCV model is a generalization of the SU(2) model with a general mass and a gyromagnetic ratio.  Very recently, the phase diagrams of this $p$-wave superconductor model have been further studied in the five-dimensional soliton and black hole backgrounds~\cite{Cai:2014ija}. However, all these holographic models were constructed only in the relativistic spacetimes.  Thus we wonder  whether the above results  still hold in nonrelativistic spacetimes, for example, the Lifshitz spacetime, which is our motivation in this paper.

 As we know,  many condensed matter systems exhibit the anisotropic scaling of spacetime being characterized by the dynamical critical exponent $z$ as $t\rightarrow b^z t, x^i\rightarrow b x^i$ ($z\neq 1$). This is the so-called Lifshitz fixed point. The authors of~\cite{Kachru} proposed the  $D=d+2$-dimensional gravity description dual to this scaling as $ds^2=L^2\left(-r^{2z}dt^2+r^2\Sigma^{i=d}_{i=1}dx_i^2+\frac{dr^2}{r^2}\right)$,
where $r\in(0,\infty)$ and $L$  is the radius of curvature. An alternative way to realize the Lifshitz gravitational spacetime is from a massless scalar field $\varpi$ coupled to an Abelian gauge field with the  action~\cite{Tylor}
\begin{equation}\label{Lifshitz action}
\mathcal{S}_g=\frac{1}{16\pi G_{d+2}}\int d^{d+2} x\sqrt{-g}\left(R-2\Lambda-\frac{1}{2}\partial_\mu\varpi\partial^\mu\varpi-\frac{1}{4}e^
{b\varpi}\mathcal{F}_{\mu\nu}\mathcal{F}^{\mu\nu}\right).
\end{equation}
The Lifshitz spacetime is generalized to a finite temperature system as~\cite{Pdw09052678}
\begin{equation}\label{Lmetric}
ds^2=L^2\left(-r^{2z}f(r)dt^2+\frac{dr^2}{r^2f(r)}+r^2\sum_{i=1}^d dx_i^2\right),
\end{equation}
where
\begin{eqnarray}
 f(r)&=&1-\frac{r_+^{z+d}}{r^{z+d}},\ \  \Lambda=-\frac{(z+d-1)(z+d)}{2L^2},\label{metric function}\\
 \mathcal{F}_{rt}&=&\sqrt{2L^2(z-1)(z+d)}r^{z+d-1},\ \ e^{b\varpi}=r^{-2d},\ \ b^2=\frac{2d}{z-1}.
\end{eqnarray}
The Hawking temperature of the black hole is given by
\begin{equation}\label{HawkingT}
T=\frac{(z+d)r_+^z}{4\pi},
\end{equation}
where $r_+$ denotes the black hole horizon. In the remainder of this paper we will set $L=1$. To see how the Lifshitz dynamical critical exponent $z$ affects the superconductor transition, it is helpful to build holographic superconductors in the Lifshitz black hole background. For related work, see, for example, Refs.~\cite{Brynjolfsson065401,Sin4617,Buyanyan,Lu:2013tza,Zhao:2013pva,Momeni:2012tw}. The scalar condensate was studied in the four-dimensional Lifshitz black hole with $z=3/2$~\cite{Brynjolfsson065401} and $z=2$~\cite{Sin4617,Buyanyan}. In particular, the author in Ref.~\cite{Buyanyan} realized the $s$-wave and $p$-wave superconductor models in the four-dimensional Lifshitz black hole~(\ref{Lmetric}). By numerical and analytical methods, we studied the holographic superconductors in four- and five-dimensional Lifshitz black hole spacetimes~(\ref{Lmetric}) in Ref.~\cite{Lu:2013tza}. It was found that as $z$ increases, the phase transition becomes difficult and the superconductivity becomes weak. However, the critical exponent for the superconductor transition is always $\frac{1}{2}$ that is independent of $z$ and the spacetime dimension. Following Ref.~\cite{Albash:2008eh}, in the probe limit, the holographic superconductor with an external magnetic field was studied analytically in Ref.~\cite{Zhao:2013pva}, and the results showed that the increasing $z$ hinders the scalar condensate and the Lifshitz scaling does not modify the well-known Ginzburg-Landau relation for the upper critical magnetic field.

On the basis of the above motivation, in this paper, we will study the Lifshitz effects  on the superconductor transition induced  by an applied magnetic field by coupling the MCV field  into the  Lifshitz spacetime (\ref{Lmetric}) on the probe approximation. The results show that in the presence of the magnetic field, the Landau level is obtained, and the vector condensate can always be triggered whether the charge density of the system vanishes or not.  Moreover, the increasing $z$ improves the response of the critical temperature $T_c$ to the applied magnetic field with the lowest Landau level. However, the increasing $z$ inhibits the phase transition in the case of the excited Landau level. In addition, we also find that even in the Lifshitz spacetime, the SU(2) Yang-Mills (YM) model is still a generalization of the  MCV model. Finally, in the Appendix, we study the condensate of the MCV model without the magnetic field. The results show that below $T_c$ the vector field begins to condense with a critical exponent $1/2$, and the increasing $z$ hinders the phase transition.

This paper is organized as follows. In Sec.~II, we study the holographic $p$-wave superconductor phase transition induced by the applied magnetic field  in the MCV model. The holographic $p$-wave phase transition in the SU(2) YM model is studied in Sec.~III. The final section is devoted to conclusions and  discussions. In the Appendix, we study  the condensate of the vector field in the absence of the magnetic field.

\section{Maxwell complex vector model in Lifshitz spacetime}

In this section, we study the holographic $p$-wave superconductor induced by an applied magnetic field in the four-dimensional Lifshitz gravity coupled to the MCV field.

Following Ref.~\cite{Cai:2013aca}, we consider the  matter action  consisting of a Maxwell field and a complex vector field,
\begin{equation}\label{Lvector}
\mathcal{S}_{MCV}=\frac{1}{16 \pi G_4}\int dx^4\sqrt{-g}\left(-\frac{1}{4}F_{\mu\nu}F^{\mu\nu}-\frac{1}{2}\rho^\dag_{\mu\nu}\rho^{\mu\nu}-m^2\rho^\dag_\mu
\rho^\mu+iq\gamma\rho_\mu\rho^\dag_\nu F^{\mu\nu}\right),
\end{equation}
where $F_{\mu\nu}=\nabla_\mu A_\nu-\nabla_\nu A_\mu$ is the strength of the U(1) gauge field $A_\mu$. The tensor $\rho_{\mu\nu}$ is defined by $\rho_{\mu\nu}=D_\mu\rho_\nu-D_\nu\rho_\mu$ with the covariant derivative $D_\mu=\nabla_\mu-iq A_\mu$, and $m$ ($q$) is the mass (charge) of the vector field $\rho_\mu$. The last term with a parameter $\gamma$ represents the interaction between the vector field $\rho_\mu$ and the gauge field $A_\mu$, which  plays the role of inducing the phase transition. Comparing this action with that in Ref.~\cite{Djukanovic:2005ag}, we ignore the neutral part of the vector meson because it does not contribute to the condensate of the  charged meson.

In this paper, we neglect the backreaction of the matter sector (\ref{Lvector}) on the
Lifshitz background~(\ref{Lmetric}). This is the so-called probe limit, which can be realized by taking $q\rightarrow\infty$ with $q\rho_\mu$ and $qA_\mu$ fixed. A variation of the action (\ref{Lvector}) with respect to the vector field $\rho_\mu$ provides us with the equation of motion,
\begin{equation}\label{EOMrho}
 D^\nu\rho_{\nu\mu}-m^2\rho_\mu+iq\gamma\rho^\nu F_{\nu\mu}=0,
 \end{equation}
 while the  equation of motion for the gauge field can be obtained by varying the action (\ref{Lvector}) with respect to  the electromagnetic potential $A_\mu$,
 \begin{equation}\label{EOMphi}
  \nabla^\nu F_{\nu\mu}-iq(\rho^\nu\rho^\dag_{\nu\mu}-\rho^{\nu\dag}\rho_{\nu\mu})
  +iq\gamma\nabla^\nu(\rho_\nu\rho^\dag_\mu -\rho^\dag_\nu\rho_\mu)=0.
\end{equation}

To model the vector condensate induced by a magnetic field, we turn on a magnetic field $B$ perpendicular to the $x-y$ plane as well as a vector field
\begin{eqnarray}
\rho_\nu dx^\nu&=&\left(\epsilon\varrho_x(r,x)e^{ipy}+\mathcal{O}(\epsilon^3)\right)dx+\left(\epsilon\varrho_y(r,x)e^{ipy}e^{i\theta}+
\mathcal{O}(\epsilon^3)\right)dy,\\
A_\nu dx^\nu&=&\left(\phi(r)+\mathcal{O}(\epsilon^2)\right) dt+\left(Bx+\mathcal{O}(\epsilon^2)\right)dy,
\end{eqnarray}
where $\epsilon$ is a small parameter characterizing the deviation from the critical point. $\varrho_x(r,x),\varrho_y(r,x)$ and $\phi(r)$ are all real functions, and $p$ is a constant, while $\theta$ is the phase difference between the $x$ and $y$ components of $\rho_\mu$. Substituting the above ansatz into Eq.~(\ref{EOMphi}), the equation of $\phi(r)$ at the linear level simplifies to
\begin{equation}
\phi''+\frac{3-z}{r}\phi'=0.
\end{equation}
According to the gauge/gravity dual dictionary, near the boundary $r\rightarrow\infty$, the leading term of the asymptotical expansion for $\phi(r)$ gives the chemical potential $\mu$ of the dual theory. To satisfy the norm of $A_\mu$ at the horizon, we impose $\phi(r_+)=0$. Therefore, the gauge field $\phi(r)$ takes the form
\begin{equation}
\phi(r)=\mu\left(1-(\frac{r_+}{r})^{2-z}\right).
\end{equation}
To seek the solution for $\varrho_x$ and $\varrho_y$, we should solve Eq.~(\ref{EOMrho}) at linear order. In order to satisfy the equation of motion with the given ansatz, the phase difference $\theta$ can only be chosen as $\theta_+=\frac{\pi}{2}+2 n\pi$ and  $\theta_-=-\frac{\pi}{2}+2 n\pi$ with an arbitrary integer $n$. Making a variable separation as $\varrho_x(r,x)=\varphi_x(r)U(x)$ and $\varrho_y(r,x)=\varphi_y(r)V(x)$, at the linear level we have
\begin{eqnarray}
\varphi_x (r) \dot{U}(x)\pm ( q B x-p)\varphi_y (r) V(x)=0,\label{eomX} \\
\varphi'_x (r) \dot{U}(x)\pm (q B x-p)\varphi'_y (r) V(x)=0,\label{eomY}
\end{eqnarray}
\begin{eqnarray}
\varphi''_x+\left(\frac{z+1}{r}+\frac{f'}{f}\right)\varphi'_x-\frac{m^2}{r^2f}\varphi_x+\frac{q^2\phi^2}{r^{2(z+1)}f^2}\varphi_x+
\ \ \ \ \ \ \ \ \ \ \ \ & &\nonumber \\
\ \ \ \ \ \ \ \ \ \ \ \frac{\varphi_x}{r^4f}\left(-(q B x-p)^2\pm (p-q B x)\frac{\dot{V}}{U}\frac{\varphi_y}{\varphi_x}\pm\frac{\gamma q B V}{U}\frac{\varphi_y}{\varphi_x}\right)&=&0, \label{eomvarpx}\\
\varphi''_y+\left(\frac{z+1}{r}+\frac{f'}{f}\right)\varphi'_y-\frac{m^2}{r^2f}\varphi_y+\frac{q^2\phi^2}{r^{2(z+1)}f^2}\varphi_y+\ \ \ \ \ \ \ \ \ \ \ \ \ &&\nonumber\\
\frac{\varphi_y}{r^4f}\left(\frac{\ddot{V}}{V}\pm (1+\gamma)q B \frac{U}{V}\frac{\varphi_x}{\varphi_y}\pm ( q B x-p )\frac{\dot{U}}{V}\frac{\varphi_x}{\varphi_y}\right)&=&0,\label{eomvarpy}
\end{eqnarray}
where the prime (dot) denotes the derivative with respect to $r$ ($x$), while the upper sign and the lower sign correspond to $\theta_+$ and $\theta_-$, respectively. To satisfy Eqs.~(\ref{eomX}) and (\ref{eomY}), we impose the constraints
\begin{equation}\label{eomC}
\varphi_y=c\varphi_x, \ \ \ \dot{U}\pm c(qBx-p)V=0,
\end{equation}
with a real constant $c$. Substituting Eq.~(\ref{eomC}) into Eqs.~(\ref{eomvarpx}) and (\ref{eomvarpy}), we have three equations
\begin{eqnarray}
-\ddot{U}\mp q c B(1+\gamma)V+(q B x-p)^2U- E U&=&0,\label{eigX}\\
-\ddot{V}\mp \frac{q B(1+\gamma)}{c}U+(q B x-p)^2V -E V&=&0, \label{eigY}\\
\varphi''_x+\left(\frac{z+1}{r}+\frac{f'}{f}\right)\varphi'_x-\frac{m^2}{r^2f}\varphi_x+\frac{q^2\phi^2}{r^{2(z+1)}f^2}
\varphi_x-\frac{E}{r^4f}\varphi_x&=&0,\label{eigvarpx}
\end{eqnarray}
where $E$ is the eigenvalue that can be obtained by solving Eqs.~(\ref{eigX}) and (\ref{eigY}). It is evident that Eqs.~(\ref{eigX}) and (\ref{eigY}) are the same as the ones in Ref.~\cite{Cai:2013pda}. We still take $c^2=1$ to make sure that the equations of $U(x)$ and $V(x)$ have exact solutions. Introducing a new function as
\begin{equation}\label{ansXY}
\psi(x)=U(x)-V(x),
\end{equation}
and subtracting Eq.~(\ref{eigX}) from Eq.~(\ref{eigY}), we get the harmonic oscillator equation
\begin{equation}\label{eompsi}
\ddot{\psi}+(E\mp q c B(1+\gamma)-(q B x-p)^2)\psi=0.
\end{equation}

Defining a new variable
$\xi=\sqrt{|q B|}(x-\frac{p}{q B})$ and constant $\eta=\frac{E\mp q c B(1+\gamma)}{|q B|}$, Eq.~(\ref{eompsi}) simplifies to
\begin{equation}\label{hermi}
\frac{d^2\psi(\xi)}{d\xi^2}+(\eta-\xi^2)\psi(\xi)=0.
\end{equation}
As we all know, the standard boundary condition for the wave function requires $\psi(\xi)$ to be finite as $|\xi|\rightarrow \infty$, from which we must choose the regular asymptotical solution $\psi(\xi)\sim e^{-\frac{\xi^2}{2}}$. Plugging $\psi(\xi)=e^{-\frac{\xi^2}{2}} H(\xi)$ into Eq.~(\ref{hermi}) yields the Hermite equation in terms of $H(\xi)$.
Expanding $H(\xi)$ near $\xi=0$ by the Taylor series, we find that only when $\eta_n=2n+1~(n=0,~1,~2,\cdots)$ is there a regular solution to the Hermite equation. Concretely, the solution $\psi(x)$ can be written in terms of the Hermite function
\begin{equation}\label{solupsi}
\psi(x)=N_n e^{-\frac{1}{2}|q B|(x-\frac{p}{q B})^2}H_n(\sqrt{|q B|}(x-\frac{p}{q B})),
\end{equation}
while the corresponding eigenvalue, i.e., the so-called Landau level, is of the form
\begin{equation}\label{Enphi}
E_n=(2n+1)|q B|\pm q c B(1+\gamma),
\end{equation}
where $N_n$ and $H_n$ denote a normalization constant, the Hermite function, respectively. From the eigenvalue~(\ref{Enphi}), we find that the nonminimal coupling $\gamma$ between the gauge field and the matter field  leads to the lowest eigenvalue, which is negative and thus will bring interesting results.

 We can read off the effective mass of $\varphi_x$ corresponding to the eigenvalue $E_n$
\begin{equation}\label{m2eff}
m^2_{eff}=m^2-\frac{q^2\phi^2}{r^{2z}f}+\frac{E_n}{r^2}=m^2-\frac{q^2\phi^2}{r^{2z}f}+\frac{(2n+1)|q B|\pm qc B(1+\gamma)}{r^2},
\end{equation}
which depends on the magnetic field $B$, especially on the nonminimal coupling parameter $\gamma$.

In the case of $n=0$, we can obtain the lowest Landau level
\begin{equation}\label{lowEn}
E^L_0=-|\gamma q B|
\end{equation}
by taking sign$(qcB)=\mp$ and $\gamma>0$, where ``$-$" (``$+$") corresponds to $\theta_+$ ($\theta_-$).
From Eq.~(\ref{m2eff}), the effective mass of $\varphi_x$ at the lowest Landau level is given by
\begin{equation}\label{effm}
m^2_{eff}=m^2-\frac{q^2\phi^2}{r^{2z}f}-\frac{|\gamma q B|}{r^2}.
\end{equation}
 It is clear that the increasing magnetic field decreases the effective mass and thus tends to enhance the superconductor phase transition, while the increasing mass  will increase $m^2_{eff}$ and make the phase transition difficult.

 Equation~(\ref{eigvarpx}) with the effective mass~(\ref{effm}) can be written concretely as
 \begin{equation}\label{eomvarpc}
\varphi_x''+\left(\frac{z+1}{r}+\frac{f'}{f}\right)\varphi_x'-\frac{m^2}{r^2f}\varphi_x+\frac{q^2\phi^2}{r^{2(z+1)}f^2}
\varphi_x+\frac{|\gamma q B|}{r^4f}\varphi_x=0.
\end{equation}
 Near the boundary $r\rightarrow\infty$, the asymptotical expansion of $\varphi_x(r)$ is of the form
\begin{equation}\label{expvarx}
\varphi_x(r)=\frac{\varphi_{x-}}{ r^{\Delta_-}}+\frac{\varphi_{x+}}{ r^{\Delta_+}}+\cdots,
\end{equation}
where $\Delta_\pm=\frac{1}{2}(z\pm\sqrt{z^2+4m^2})$. Since the coefficient $\varphi_{x-}$ is dominant near the boundary, according to the AdS/CFT dictionary,  we interpret $\varphi_{x-}$  as the source of the $x$ component of the dual vector operator $J_x$ in the boundary field theory, and $\varphi_{x+}$ as the vacuum expectation value of $J_x$.  In order to meet the requirement that the U(1) symmetry should be broken spontaneously, we impose  the source-free condition, i.e., $\varphi_{x-}=0$.

Now, let us consider the simple case with the vanishing electric field $\phi(r)$. By changing to a dimensionless variable $u=\frac{r_+}{r}$, Eq.~(\ref{eomvarpc})  can be written as
\begin{equation}\label{eomzeta}
\varphi_x''-\frac{z-1+3u^{z+2}}{u(1-u^{z+2})}\varphi_x'-\frac{m^2}{u^2(1-u^{z+2})}\varphi_x+(\frac{z+2}
{4\pi})^{\frac{2}{z}}\frac{\zeta }{1-u^{z+2}}\varphi_x=0,
\end{equation}
with $\zeta=|\gamma q B|/T^{2/z}$, where we have used the temperature~(\ref{HawkingT}).  Near the critical point, we will encounter a marginally stable mode corresponding to the solution of Eq.~(\ref{eomzeta}). To solve such the equation, we should impose the boundary conditions, i.e., the regular condition at the horizon, as well as the source-free condition at infinity $\varphi_{x-}=0$. For a given  $m^2$ and  $z$, only certain special values of $\zeta$ can satisfy the equation. Concretely, we first solve this equation by using the shooting method.

\begin{figure}
\includegraphics[width=2.9in]{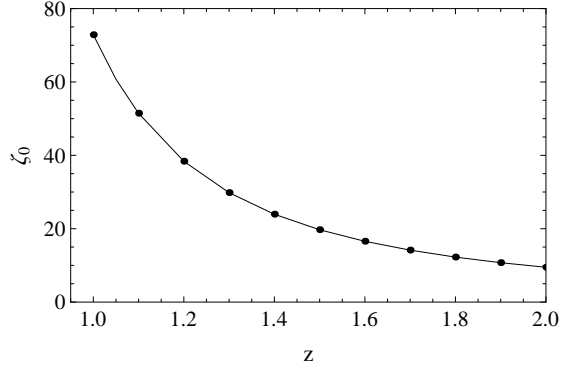}
\caption{The critical value $\zeta_0$ versus the Lifshitz dynamical exponent $z$ of the vector field $\rho_\mu$ for $\Delta_+=3/2$. The black points are obtained by using the shooting method to solve Eq.~(\ref{eomzeta}).}
\label{vanqwz}
\end{figure}

We plot the critical value $\zeta_0$ as a function of $z$ with fixed $\Delta_+=\frac{3}{2}$ in Fig.~\ref{vanqwz}. Before analyzing the Lifshitz effect on $\zeta_0$, it is necessary to determine which side of the phase boundary is the superconducting phase. For a given Lifshitz exponent $z$ and the applied magnetic field $B$, when we decrease the temperature $T$, the normal phase will become instable to getting across the critical point; then, it will enter into the condensed phase, which corresponds to increasing $\zeta=|\gamma q B|/T^{2/z}$. Therefore, the upper right region in the figure stands for the condensed phase while the other region represents the normal phase. From the figure, we find that $\zeta_0$ decreases with the increasing $z$, which indicates that the increasing $z$ enhances the phase transition for the fixed $|\gamma q B|$.

\begin{figure}
\includegraphics[width=2.9in]{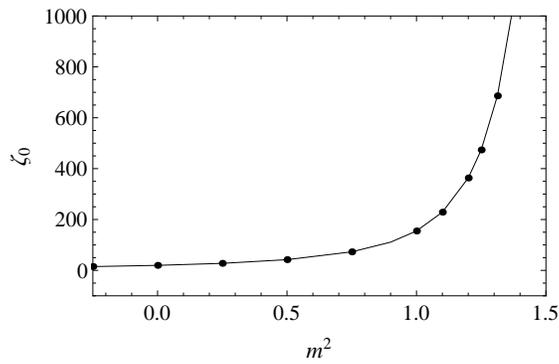}
\caption{The critical value $\zeta_0$ with respect to the mass squared $m^2$ of the vector field $\rho_\mu$ for $\Delta_+=3/2$. The black points are obtained by using the shooting method to solve Eq.~(\ref{eomzeta}). }
\label{vanqwm}
\end{figure}

In addition, we show the critical value $\zeta_0$ versus the mass squared $m^2$ of the vector field with fixed $\Delta_+=\frac{3}{2}$ in Fig.~\ref{vanqwm}. According to the analysis of $\zeta=|\gamma q B|/T^{2/z}$, we can easily estimate that the upper left region in the figure denotes the condensed phase. From the figure, it is clear that $\zeta_0$ improves with the increase of $m^2$. Therefore, we can conclude that the increasing $m^2$ inhibits the superconductor phase transition, which qualitatively agrees with in Ref.~\cite{Cai:2013pda}.

 To uphold the above numerical results, we then solve Eq.~(\ref{eomzeta}) by using the alternative analytical method, i.e., the SL eigenvalue method~\cite{Siopsis}. By introducing a trial function $\Gamma(u)$ related to $\varphi_x(u)$ as
\begin{equation}
\varphi_x(u)=\langle J_x\rangle {\left(\frac{u}{r_+}\right)}^{\Delta_{+}}\Gamma(u),
\end{equation}
 the equation of motion for $\Gamma(u)$ is given by
\begin{eqnarray}
\Gamma''(u)+\frac{(\sqrt{4 m^2+z^2}+z+3)u^{z+2}-\sqrt{4 m^2+z^2}-1}{u \left(u^{z+2}-1\right)}\Gamma'(u)\ \ \  \ \ \ \ \ \ \ \ \ \ \ \ \ \ \  &&\nonumber \\
-\frac{\frac{1}{2}(z+2)(\sqrt{4 m^2+z^2}+z) u^z- \zeta(\frac{z+2}{4\pi})^{\frac{2}{z}}+m^2 u^z}{1-u^{z+2}}\Gamma(u)&=&0,
\end{eqnarray}
with the boundary conditions $\Gamma(0)=1$ and $\Gamma'(0)=0$.  Such an equation can be further written as the SL
eigenvalue equation
\begin{eqnarray}
&&\frac{d}{du}\big(\underbrace{(1-u^{z+2}) u^{\sqrt{4 m^2+z^2}+1}}_{K}\Gamma '(u)\big)-\underbrace{\frac{1}{2} \left((z+2) (\sqrt{4m^2+z^2}+z)+2 m^2\right) u^{\sqrt{4 m^2+z^2}+z+1}}_P\Gamma(u)\nonumber \\
&&\qquad \qquad \qquad \qquad \qquad \qquad\qquad \qquad \qquad \qquad +\zeta  \underbrace{\left(\frac{z+2}{4\pi}\right)^{\frac{2}{z}}u^{\sqrt{4 m^2+z^2}+1}}_Q\Gamma(u)=0.
\end{eqnarray}
The minimum eigenvalue of $\zeta$ can be obtained by varying the following function
\begin{equation}\label{integ}
\zeta=\frac{\int^1_0du(K{\Gamma'}^2+P\Gamma^2)}{\int^1_0duQ\Gamma^2}.
\end{equation}
To estimate the eigenvalue, we take the trial function $\Gamma(u,\alpha)=1-\alpha u^2$, with the constant $\alpha$ to be determined. Then we can obtain the minimum value of $\zeta$ from Eq.~(\ref{integ}) for the given $m$ and $z$. We list the results from the shooting method and the SL method in Table~\ref{tab:resofzeta} for a clear comparison.
\begin{table}
\caption{ The critical value of $\zeta=|\gamma q B|/T^{2/z}$ obtained  by using shooting method and the SL eigenvalue method. For all cases, we fix $\Delta_+=3/2$.}\label{tab:resofzeta}
\begin{ruledtabular}
\begin{tabular}{c  c c c  c c c}
  &$z=1$&$z=6/5$&$z=7/5$&$z=8/5$&$z=9/5$&$z=2$ \\ \hline
  \text{Shooting}&72.842&38.371&23.941&16.569&12.261&9.490 \\
   \text{SL}&72.875&38.396&23.962&16.587&12.278&9.506 \\ \hline\hline
   &$m^2=-1/4$&$m^2=1/4$&$m^2=1/2$&$m^2=1$&$m^2=5/4$&$m^2=3/2$\\ \hline
    \text{Shooting}&14.895&27.629&42.096&154.872&473.552&3006.819 \\
  \text{SL}&14.912&27.651&42.122&154.921&473.649&3007.164\\
  \end{tabular}
\end{ruledtabular}
\end{table}
It follows that the analytical results are in good agreement with the numerical results. In particular, when $z=1$, the results are the same as the ones in Ref.~\cite{Cai:2013pda}. Therefore, we can conclude that the analytical method is powerful for this MCV model.

Note that in the literature, for example, Refs.~\cite{Hartnoll2008,Buyanyan}, the applied magnetic field was turned off. When the temperature decreases to a critical value, the gauge field $A_\mu$ has nonvanishing mass,  which results in the spontaneous breaking of the U(1) symmetry, going with the condensate. Hence, the matter field will not condense if we turn off the electric field. However, in the presence of the applied magnetic field, even though the electric field vanishes, the instability of the black hole can still be triggered. This interesting result is similar to the  QCD vacuum instability, which is  induced by the strong magnetic field and develops  the condensate of the $\rho$ meson~\cite{Chernodub:2010qx,Chernodub:2011mc}. From the action (\ref{Lvector}), we can clearly see  it is the nonminimal coupling term between the vector field $\rho_\mu$ and the U(1) gauge field $A_\mu$ that leads to the phase transition.

To systemically study the effect of  $z$ on the critical value in the system with the applied magnetic field $B$, next we consider the case with the fixed charge density $\rho$ and the general Landau level $E_n$. The equation of motion for $\varphi_x(u)$ reads
\begin{equation}
\varphi_x''-\frac{z-1+3u^{2+z}}{u(1-u^{2+z})}\varphi_x'-\frac{m^2}{u^2(1-u^{2+z})}\varphi_x+\frac{q^2(u^2-u^z)^2
\lambda^2}{u^2(1-u^{2+z})^2}\varphi_x-\frac{E_n \lambda}{(1-u^{2+z})\rho}\varphi_x=0,
\end{equation}
where $\lambda=\frac{\rho}{r_+^2}$, and $E_n$ is of the form (\ref{Enphi}). For convenience, we introduce a new function $F(u)$
\begin{equation}
\varphi_x(u)=(\frac{u}{r_+})^{\Delta_-}F(u).
\end{equation}
Then we get
\begin{eqnarray}
F''+ \left(\frac{2 \Delta_-}{u}+\frac{3 u^{z+2}+z-1}{u( u^{z+2}-1)}\right)F'+\frac{ m+\Delta_- \left(z-\Delta_- +(\Delta_- +2) u^{z+2}\right)}{u^2\left(u^{z+2}-1\right)}F&&\nonumber \\
+\left(\frac{\lambda ^2 q^2\left(u^2-u^z\right)^2}{u^{2}\left(u^{z+2}-1\right)^2}+\frac{E_n  \lambda  }{\rho \left(u^{z+2}-1\right)}\right)F&=&0.
\end{eqnarray}
 It is easy to see that this equation depends on two dimensionless  parameters, i.e., $E_n/\rho$ and $\lambda$. Under the regular condition at the horizon ($u=1$) and the source-free condition near the boundary ($u=0$), only if these two parameters satisfy a certain relation does the equation have the nontrivial solution.

\begin{figure}
  \includegraphics[width=3.2 in]{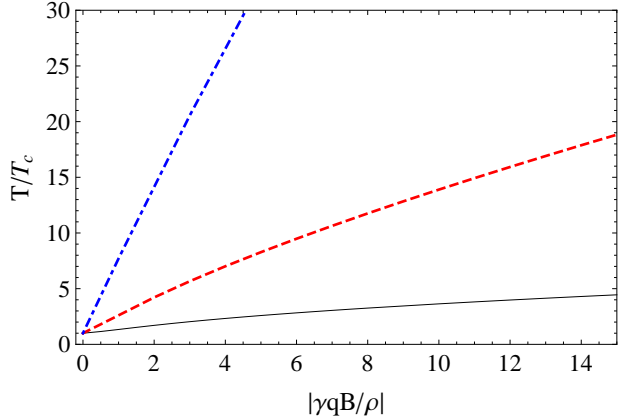}
  \caption{ The critical temperature versus the magnetic field with different $z$ and the fixed $\Delta_+=3/2$ for the lowest Landau level ($E_0^L=-|\gamma q B|$). The lines from bottom to top correspond to $z=1$~(black  solid), 3/2~(red  dashed), 9/5~(blue  dot-dashed), respectively.}
\label{nvanqnb}
\end{figure}

 We still consider the case of the lowest Landau level ($E_0^L=-|\gamma q B|$) with fixed $\Delta_+=3/2$. The critical temperature $T_c$ as a function of the magnetic field $B$ with different $z$ is plotted in Fig.~\ref{nvanqnb}, from which we have the following comments: for the fixed $z$, when $|\gamma q B/\rho|$ increases, $T_c$ increases; it is obvious from the effective mass that the increasing $B$ decreases $m_{eff}^2$ and thus raises the critical temperature. For the fixed $|\gamma q B/\rho|$, when $z$ increases~($z=1,~3/2,~9/5$), the ratio $T/T_c$ increases, which means that the effect of the external magnetic field on $T_c$ becomes more obvious. So far, we can summarize that in the case of the lowest Landau level, the increasing $z$ improves the response of $T_c$ to the applied magnetic field regardless of the charge density. Figure~\ref{nvanqnb} is similar to the one in Ref.~\cite{Callebaut:2013ria} where the chiral critical temperature improves with the increase of the applied magnetic field.

\begin{figure}
  \includegraphics[width=3.2 in]{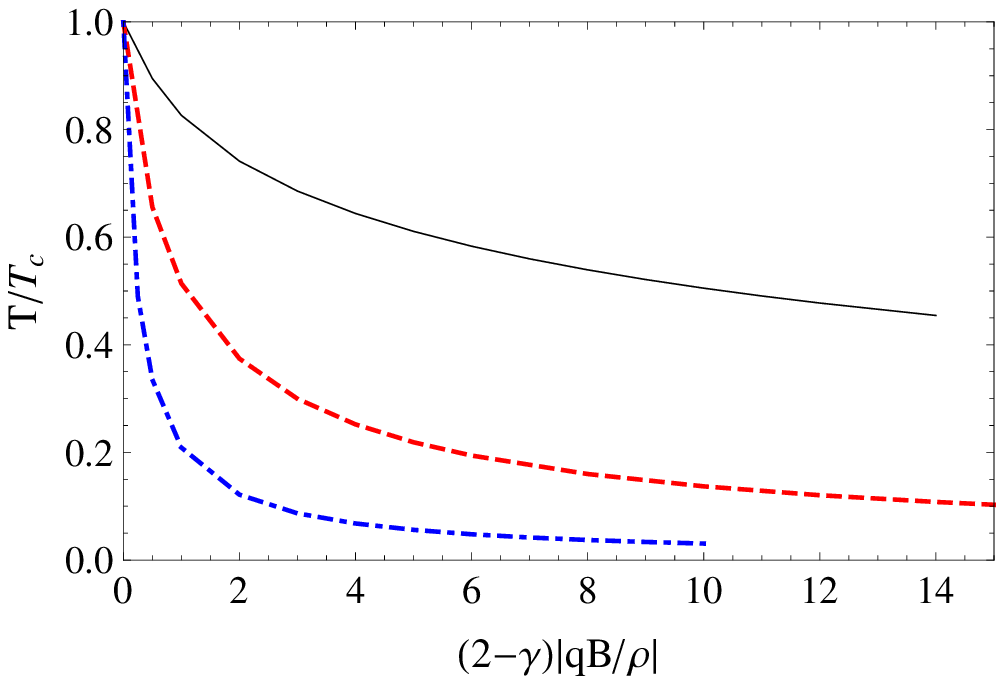}
  \caption{The critical temperature as a function of the magnetic field with different $z$ and the fixed $\Delta_+=3/2$ for the excited Landau level ($E_1=(2-\gamma)| q B|$). The lines from top to bottom correspond to $z=1$~(black  solid), 3/2~(red  dashed), 9/5~(blue  dot-dashed), respectively.}
\label{nvanqpb}
\end{figure}

 To compare with the case of the lowest Landau level $E_0^L=-|\gamma q B|$ in Fig.~\ref{nvanqnb}, we plot the ratio $T/T_c$ versus $|q B/\rho|$ for the excited Landau level $E_1=(2-\gamma)|q B|$ in Fig.~\ref{nvanqpb}, which can be obtained by taking sign$(qcB)=\mp$, $\gamma>0$ and $n=1$ from Eq.~(\ref{Enphi}). It should be noted that when $E_1<0$, corresponding to the case of $\gamma>2$, the effect of the magnetic field on the superconductor phase transition is similar to the one in the case of the lowest Landau level. Therefore, in order to qualitatively illustrate the difference between the excited Landau level with $E_n>0$ and the lowest Landau level with $E_0<0$, here we have chosen $E_1>0$ corresponding to the case of $0<\gamma<2$, which can reflect the different effect of the magnetic field on the superconductor phase transition from that of the lowest Landau level $E_0$.
It follows that the increasing magnetic field hinders the phase transition, which is the common property of the ordinary superconductor~\cite{Albash:2008eh,Nakano:2008xc}. Moreover, the fact that the increasing $z$ decreases the critical temperature means that in the case of the excited Landau level, the increasing $z$ hinders the conductor/superconductor phase transition, which is similar to the Lifshitz effect on the MCV model with only the scalar potential $A_t$ turned on. See the Appendix for the detailed calculation of $T_c$ about the MCV model without the magnetic field.

\section{Yang-Mills model in Lifshitz spacetime}
The author of Ref.~\cite{Wong:2013rda} discussed the holographic superconductor phase transition induced by the non-Abelian magnetic field in the black hole background. To compare our MCV model with the non-Abelian model, in this section, we study the $p$-wave superconductor phase transition triggered by the magnetic field in the  Lifshitz black hole coupled to the SU(2) field in the probe limit. The SU(2) YM action is given by~\cite{Gubser2008a}
\begin{equation}\label{Lpwave}
  \mathcal{S}_{YM}=\frac{1}{16\pi G_4}\int dx^4\sqrt{-g}\left(-\frac{1}{4}F^a_{\mu\nu}F^{a\mu\nu}\right),
\end{equation}
where $F^a_{\mu\nu}=\partial_\mu A^a_\nu-\partial_\nu A^a_\mu+\varepsilon^{abc}A^b_\mu A^c_\nu$ is the field strength of the gauge field with the gauge index ($a,~b,~c=1,~2,~3$). The SU(2) group has three generators which satisfy the commutation relation $[\tau^i,\tau^j]=\varepsilon^{ijk}\tau^k$ ($i,~j,~k=1,~2,~3$).   The equation of motion of the gauge field $A=A^a_\mu\tau^a dx^\mu$ reads
\begin{equation}\label{Eomp}
   \nabla_\mu F^{a\mu\nu}+\varepsilon^{abc}A^b_\mu F^{c\mu\nu}=0.
\end{equation}
Concretely, we take the ansatz for the gauge field as
\begin{eqnarray}\label{Su2ans}
A^1_\mu dx^\mu&=&\left(\epsilon a^1_x(r,x,y)+\mathcal{O}(\epsilon^3)\right) dx+\left(\epsilon a^1_y(r,x,y)+\mathcal{O}(\epsilon^3)\right)dy,\nonumber\\
A^2_\mu dx^\mu&=&\left(\epsilon a^2_x(r,x,y)+\mathcal{O}(\epsilon^3)\right)dx+\left(\epsilon a^2_y(r,x,y)+\mathcal{O}(\epsilon^3)\right)dy,\\
A^3_\mu dx^\mu&=&\left(\phi(r)+\mathcal{O}(\epsilon^2)\right)dt+\left(B x+\mathcal{O}(\epsilon^2)\right)dy,\nonumber
\end{eqnarray}
where $\epsilon$ is a small parameter characterizing the deviation from the critical point.

Substituting the ansatz (\ref{Su2ans}) into Eq.~(\ref{Eomp}), we can read off the equations of motion
\begin{eqnarray}
\partial_x W_x+(\partial_y-i B x)W_y&=&0,\\
\partial_x\partial_r W_x+(\partial_y-i B x)\partial_r W_y&=&0,
\end{eqnarray}
\begin{eqnarray}
\partial_r^2W_x+\left(\frac{z+1}{r}+\frac{f'}{f}\right)\partial_r W_x
+\frac{1}{r^4f}\bigg((\partial^2_y-2 i B x \partial_y-B^2x^2+\frac{\phi^2}{r^{2z-2}f})W_x+&&\nonumber\\
\qquad \qquad  \qquad \qquad \qquad  \qquad \qquad \qquad  \qquad (-\partial_x\partial_y+i B x\partial_x-i B)W_y\bigg)&=&0,\\
\partial_r^2W_y+\left(\frac{z+1}{r}+\frac{f'}{f}\right)\partial_r W_y
+\frac{1}{r^4f}\bigg((\partial_x\partial_y-i  B x \partial_x+ 2 i B)W_x+&&\nonumber\\
 (\partial_x^2+\frac{\phi^2}{r^{2z-2}f})W_y\bigg)&=&0,
\end{eqnarray}
where we have defined $W_x=a^1_x-ia^2_x$ and $W_y=a^1_y-ia^2_y$. To solve the above four equations, we further take a separable form for $W_x$ and $W_y$,
\begin{equation}
W_x(r,x,y)=\tilde{\varphi}_x(r)\tilde{U}(x)e^{ipy}, \ \ \ W_y(r,x,y)=\tilde{\varphi}_y(r)\tilde{V}(x)e^{ipy}e^{i \theta},
\end{equation}
which further yields the equations of motion
\begin{eqnarray}
\tilde{\varphi}_x (r) \dot{\tilde{U}}(x)\pm (B x-p)\tilde{\varphi}_y (r) \tilde{V}(x)=0, \label{tilU}\\
\tilde{\varphi}'_x (r) \dot{\tilde{U}}(x)\pm (B x-p)\tilde{\varphi}'_y (r) \tilde{V}(x)=0,\label{tilV}
\end{eqnarray}
\begin{eqnarray}
\tilde{\varphi}''_x+\left(\frac{z+1}{r}+\frac{f'}{f}\right)\tilde{\varphi}'_x+\frac{\phi^2\tilde{\varphi}_x}{r^{2(z+1)}f^2}+
\frac{\tilde{\varphi}_x}{r^4f}\bigg(-(B x-p)^2\pm &&\nonumber \\
(p-B x)\frac{\dot{\tilde{V}}}{\tilde{U}}\frac{\tilde{\varphi}_y}{\tilde{\varphi}_x}
\pm\frac{B \tilde{V}}{\tilde{U}}\frac{\tilde{\varphi}_y}{\tilde{\varphi}_x}\bigg)&=&0,\label{tilvax} \\
\tilde{\varphi}''_y+\left(\frac{z+1}{r}+\frac{f'}{f}\right)\tilde{\varphi}'_y+
\frac{\phi^2\tilde{\varphi}_y}{r^{2(z+1)}f^2}
+\frac{\tilde{\varphi}_y}{r^4f}\bigg(\frac{\ddot{\tilde{V}}}{\tilde{V}}\pm 2B\frac{\tilde{U}}{\tilde{V}}\frac{\tilde{\varphi}_x}{\tilde{\varphi}_y}
\pm (B x-p )\frac{\dot{\tilde{U}}}{\tilde{V}}\frac{\tilde{\varphi}_x}{\tilde{\varphi}_y}\bigg)&=&0,\label{tilvay}
\end{eqnarray}
where the dot and the prime denote the derivative with respect to $x$ and $r$, respectively, while the upper sign and the lower sign correspond to the phase difference $\theta_+=\frac{\pi}{2}$ and $\theta_-=-\frac{\pi}{2}$, respectively. As we all know, for the MCV field model in the standard Schwarzschild AdS black hole and soliton backgrounds~\cite{Cai:2013pda,Cai:2013kaa}, in the case of $c^2 =1$, the SU(2)  model can be understood as a generalization of the MCV model with the parameters $m^2=0,~q=1$, and $\gamma=1$. By comparing Eqs.~(\ref{tilU})-(\ref{tilvay}) with Eqs.~(\ref{eomX})-(\ref{eomvarpy}), it is easy to see that even in the anisotropy Lifshitz background, such an SU(2) model is still a special case of the MCV model with the parameters chosen as $m^2=0,~q=1$, and $\gamma=1$, which shows that the convention of $c^2=1$ in this paper is still reasonable.

\begin{figure}
  \includegraphics[width=3.2 in]{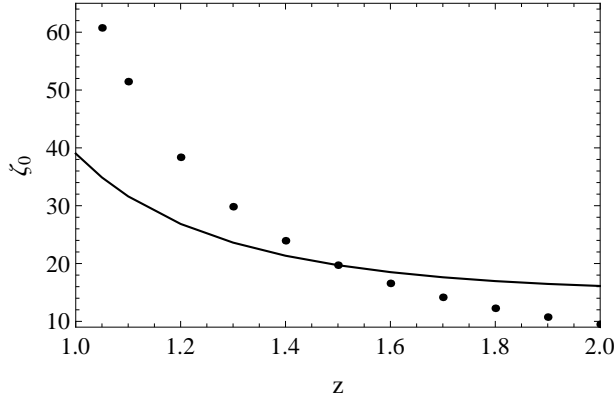}\\
  \caption{The critical value $\zeta_0$ with respect to the Lifshitz dynamical exponent $z$. The black points denote the results from the $\rho_\mu$ vector field for comparison.}
  \label{Su2nEnvc}
\end{figure}

From Eq.~(\ref{Enphi}) and the constraints ($m^2=0,~q=1$, and $\gamma=1$), we can obtain the Landau level for the SU(2) YM model, i.e., $E_n=(2n+1)|B|\pm 2 c B$, with a non-negative integer $n$. By taking  sign($cB$)$=\mp$ and $n=0$, we can get the lowest Landau level $E^L_0=-|B|$, and the corresponding effective mass of $\tilde{\varphi}_x(r)$ in the presence of the applied magnetic field and the charge density is given by
\begin{equation}
m^2_{eff}=-\frac{\phi^2}{r^{2z}f}+\frac{E_n}{r^2}=-\frac{\phi^2}{r^{2z}f}-\frac{|B|}{r^2}.
\end{equation}
Following the approaches used in Sec.~II, we first consider the simple case with vanishing charge density via the shooting method as well as the SL eigenvalue method. By using complicated calculations, we list the results in Table~\ref{tab:zetasu2} and plot the critical value $\zeta_0$ versus $z$ in Fig.~\ref{Su2nEnvc}.
\begin{table}
\caption{ The critical value of $\zeta=|B|/T^{2/z}$ calculated  by  using the shooting method and the SL eigenvalue method.}\label{tab:zetasu2}
\begin{ruledtabular}
\begin{tabular}{c  c c c  c c c}
  &$z=1$&$z=6/5$&$z=7/5$&$z=8/5$&$z=9/5$&$z=2$ \\ \hline
  \text{Shooting}&39.012&26.846&21.350&18.514&16.963&16.120 \\
   \text{SL}&39.031&26.864&21.368&18.534&16.985&16.146
  \end{tabular}
\end{ruledtabular}
\end{table}
It is easy to see that $\zeta_0$ decreases with the increasing  $z$; i.e., the increasing $z$ enhances the phase transition, which is similar to the case of the MCV model. Moreover, the analytical results agree with the numerical results. Furthermore, the solid line from the non-Abelian magnetic field intersects with the dotted line from the MCV field at the value $z=3/2$, which can be understood as follows: as we all know, in the dual field theory, the power exponent $\Delta$ of general falloff for the vector field can be regarded as the ``mass" in the field theory. In the MCV field theory, we plot $\zeta_0$ as a function of $z$ with fixed $\Delta_+=3/2$ by adjusting the mass of the vector field. However, in the SU(2) model with vanishing mass, the power exponent $\Delta_+=z$ varies with $z$. When $z=3/2$, the exponent of the vector operator in the SU(2) field model is the same as that in the MCV field model; therefore, the critical value from the two models is identical. When $z<3/2$, the ``mass" of the SU(2) field is less than that in the MCV field, so the critical temperature in the SU(2) field system is larger than that in the latter system.

We also calculate the critical temperature $T_c$ in the SU(2) model with the charge density $\rho$ and the lowest Landau level $E_0^L=-|B|$.
 \begin{figure}
  \includegraphics[width=3.2 in]{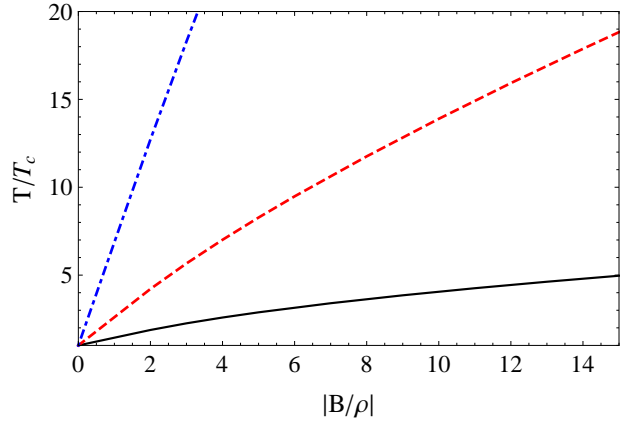}\\
  \caption{The critical temperature as a function of the magnetic field with different $z$  for the lowest Landau level ($E_0^L=-|B|$). The lines from bottom to top correspond to $z=1$~(black  solid), 3/2~(red  dashed), 9/5~(blue dot-dashed), respectively.}
  \label{Su2nEnchar}
\end{figure}
The value of $T/T_c$ as a function of the magnetic field is plotted in Fig.~\ref{Su2nEnchar}, from which we can see that for the fixed $z$, $T/T_c$ improves with the increase of $B$. For the fixed magnetic field, when $z$ increases, $T/T_c$ also increases, which means that the larger $z$ makes the phase transition easier. These results are very similar to the case of the MCV model. In particular, when $z=3/2$, the curve from the SU(2) model overlaps with the one from the MCV model, which further proves that the SU(2) field is a generalization of the MCV model. It is obvious from Eq.~(\ref{expvarx}) that $\Delta_+=3/2$ and $z=3/2$ will result in $m^2=0$. The other two cases ($z=1,~9/5$) have slight differences from the case of the MCV field ($\Delta_+=3/2$) due to the different $\Delta_+=z$ of the vector operator in the SU(2) model.

\begin{figure}
  \includegraphics[width=3.2 in]{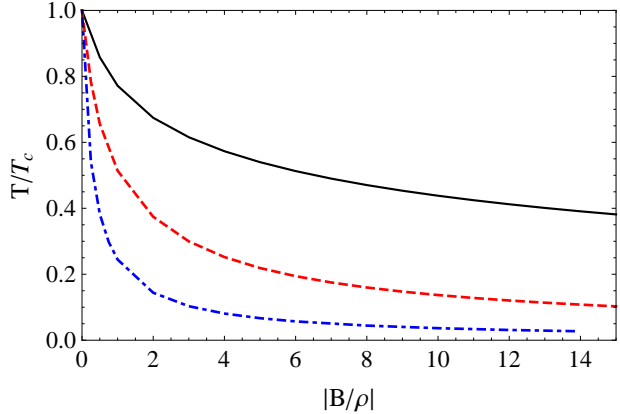}\\
  \caption{The critical temperature as a function of the magnetic field with different $z$  for the excited Landau level ($E_1=|B|$). The lines from top to bottom correspond to $z=1$~(black  solid), 3/2~(red  dashed), 9/5~(blue   dot-dashed), respectively.}
  \label{Su2pEnchar}
\end{figure}

In addition, we can obtain the excited Landau level $E_1=|B|$ by taking sign$(cB)=\mp$ and $n=1$. To compare with the case of the lowest Landau level $E^L_0=-|B|$, we plot the ratio $T/T_c$ versus  $|B/\rho|$ with the eigenvalue $E_1=|B|$ in Fig.~\ref{Su2pEnchar},
from which we can see that the critical temperature  decreases with the increasing magnetic field. Besides, the larger the Lifshitz exponent $z$, the smaller the critical temperature, which implies that the increasing $z$ makes the phase transition more difficult. In particular, the case of $z=3/2$ in Fig.~\ref{Su2pEnchar} is still identical with the case of the MCV model with $z=3/2$ in Fig.~\ref{nvanqpb}. All these results are similar to the ones of the ordinary superconductor, such as in Refs.~\cite{Albash:2008eh,Nakano:2008xc,Cai:2013pda,Zhao:2013pva}.

\section{Vortex lattice solution}
Following Ref.~\cite{Maeda:2009vf}, we will construct the vortex lattice for the MCV field by superposing the droplet solution in this section. Typically, we only consider the droplet solution with the lowest Landau level~($n=0$).

Combining Eqs.~(\ref{eomC}), (\ref{ansXY}) and (\ref{solupsi}), we get the exact solution for $U(x)$ and $V(x)$. The eigenfunction with the lowest Landau level (\ref{lowEn}) is
\begin{equation}\label{eigenXY}
U^L_0(x;p)=\frac{N_0}{2}e^{-\frac{|qB|}{2}(x-\frac{p}{qB})^2}=-V^L_0(x;p).
\end{equation}
Since the eigenvalue does not depend on the constant $p$, the linear superposition of the solutions $e^{ipy}\varphi_{xn}(r)U_n(x;p)$ and $e^{ipy}\varphi_{yn}(r)V_n(x;p)$ is still the solution of the MCV model at linear order $\mathcal{O}(\epsilon)$. To obtain the vortex lattice solution from the signal droplet solution (\ref{eigenXY}), we consider the following superposition
\begin{eqnarray}\label{latsolu}
R_l(r,x,y)&=& \varphi_{x0}(r)\sum^{+\infty}_{l=-\infty}C_le^{ip_ly}U_0^L(x;p_l)-ce^{-i\theta_\pm}\varphi_{y0}(r)e^{i\theta_\pm}\sum^{+\infty}_{
l=-\infty}C_le^{ip_ly}V_0^L(x;p_l)\nonumber\\
&=&\varphi_{x0}(r)\sum^{+\infty}_{l=-\infty}C_le^{ip_ly}\psi_0(x;p_l),
\end{eqnarray}
with $C_l=e^{-i\pi a_2 l^2/a_1^2}$ and $p_l=2\pi \sqrt{|qB|}l/a_1$, where $a_1$ and $a_2$ are arbitrary constants, and we have also used the convention $c^2=1$ and the definition (\ref{ansXY}).  Comparing the vortex lattice solution (\ref{latsolu}) with the elliptic theta function~\cite{Maeda:2009vf}, we see that the vortex lattice solution $R_l$ has two properties. The first property is  the pseudoperiodicity of the solution
\begin{eqnarray}
R_l(r,x,y)&=&R_l(r,x,y+\frac{a_1}{\sqrt{|qB|}}),\nonumber\\
R_l(r,x+\frac{2\pi}{a_1\sqrt{|qB|}},y+\frac{a_2}{a_1\sqrt{|qB|}})&=&e^{\frac{2\pi i}{a_1}(\sqrt{|qB|}y+\frac{a_2}{2a_1})}R_l(r,x,y).
\end{eqnarray}
The other property is that the cores (or the zeros) of the vortices are
located at
$\textbf{x}_{m,n}=(m+\frac{1}{2})\textbf{b}_1+(n+\frac{1}{2})\textbf{b}_2$,
where the two vectors
$\textbf{b}_1=\frac{a_1}{\sqrt{|qB|}}\partial_y$ and
$\textbf{b}_2=\frac{2\pi}{a_1\sqrt{|qB|}}\partial_x+\frac{a_2}{a_1\sqrt{|qB|}}\partial_y$,
while $m$ and $n$ are two integers. Since the vacuum expectation value of
the vector operator $J_\mu$  is proportional to the subleading
coefficient of the asymptotical expansion of $\rho_\mu$, the
combination $\langle J_\pm\rangle=\langle J_x\pm ic J_y\rangle$
shows the vortex lattice structure, which is the same as that in
Ref.~\cite{Cai:2013pda}. In addition, as for the MCV model discussed in Sec.~II, to get the lowest Landau level, we need sign$(qcB)=\mp$ and
$\gamma>0$, as well as $n=0$ with the sign ``$\mp$" corresponding
to $\theta_\pm$,  which means that in the case of $c>0$,
$\langle J_\pm\rangle$ corresponding to sign$(qB)=\mp$
 can represent the lowest Landau level, and for the case of $c<0$, so can $\langle J_\mp\rangle$ corresponding to sign$(qB)=\pm$. For both of $\theta_+$ and $\theta_-$ at the lowest Landau level, $\langle J_+\rangle$ corresponds to $qB<0$, while $\langle J_-\rangle$ corresponds to $qB>0$. In particular, if we further assume $B>0$, the vortex lattice solution of the SU(2) model will correspond to $\langle J_-\rangle$, as it is the special case of the MCV field with $q=1$ and $\gamma=1$.

\begin{figure}
  \includegraphics[width=3.2 in]{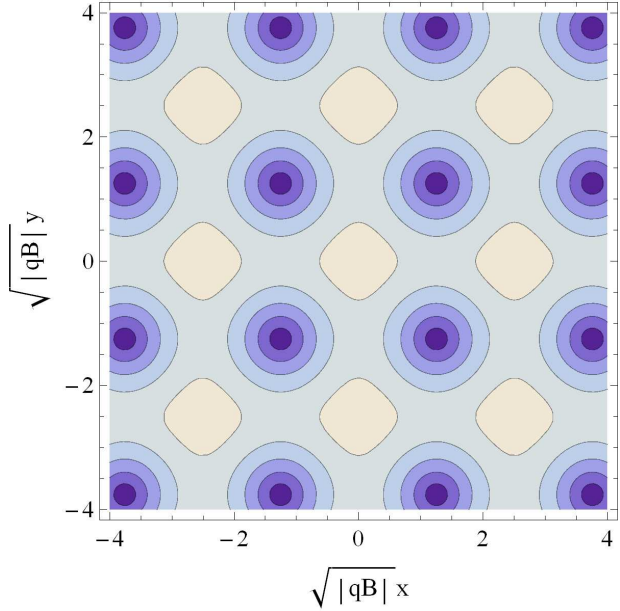}\\
  \caption{The contour plot of the square vortex lattice in the $x-y$ plane. Each darker region  denotes the core of the vortex, where the lattice solution vanishes.}
  \label{MCVVort}
\end{figure}

The triangular lattice with the parameters $a_1=2\sqrt{\pi}/3^{1/4}, a_2=2\pi/\sqrt{3}$ is shown in Ref.~\cite{Cai:2013pda}.  For the vector field condensate specializing in a spatial direction, the square lattice is also possible.  By choosing the parameters $a_1=\sqrt{2\pi}$ and $a_2=1/10000$,  the configuration of the condensate $\langle J_-\rangle$ in the $x-y$ plane for the square lattice is plotted in Fig.~\ref{MCVVort}.
If we choose other combinations of the parameters for $a_1$ and $a_2$, the different structure will be exhibited. However, to determine which case is the true ground state of the system, we should calculate the free energy of the system which includes the nonlinear effects of the holographic superconductor. By minimizing the free energy, one can obtain the combination of parameters for $a_1$ and $a_2$ of the ground state as discussed in Ref.~\cite{Maeda:2009vf}.
\section{Conclusions and discussions}
So far, in the probe limit, we have studied the properties of the holographic vector condensate induced by an external magnetic field in the four-dimensional Lifshitz black hole background by using numerical and analytical methods. Not only have we discussed the $p$-wave superconductor phase transition by introducing the MCV field and the SU(2) gauge field in the bulk, respectively, but we also emphasized the influence of the dynamical critical exponent $z$ on the critical value. The main conclusions can be summarized as follows.

For the MCV model, we found that the vector condensate can be induced by the applied magnetic field even without the charge density, and the Landau level is independent of $z$. In particular, in the case of the lowest Landau level $E_0^L=-|\gamma qB|$, as $z$ increases, the response of the ratio $T/T_c$ to the applied magnetic field becomes more obvious, which means that the increasing $z$ enhances the superconductor phase transition.
However, in the case of the excited Landau level, for example, $E_1=(2-\gamma)|qB|$, the results are opposite from the case of  $E_0^L=-|\gamma q B|$; i.e., the increasing $z$ (and the increasing magnetic field $B$) hinders the phase transition. For the SU(2) YM model, we found that even in the Lifshitz spacetime, the MCV field is still a generalization of the SU(2) model with the general mass $m$, the charge $q$, and the gyromagnetic ratio $\gamma$. Because of the diamagnetic and Pauli pair breaking effect of the magnetic field, the results in the case of $E_1=(2-\gamma)|qB|$ model the ordinary superconductors, while the results for $E_0^L=-|\gamma qB|$ are quite similar to the case of QCD vacuum instability induced by a strong magnetic field to spontaneously develop the $\rho$ meson condensate~\cite{Chernodub:2010qx,Chernodub:2011mc}. It is worth noting that some studies~\cite{Levy,Uji,Rasolt:1992zz,Klimenko:2012qi} suggested that the magnetic field can give rise to the superconductor phase transition.

In the Appendix, we also discuss the vector condensate without the applied magnetic field. Working in the probe limit, we found that near $T_c$, the vector field starts to condense via a  second-order phase transition.  Moreover, when $z$ increases, the critical temperature decreases, which means that the increasing $z$ hinders the superconductor phase transition.

It should be stressed that we focused our study of the Lifshitz black hole background on the probe approximation. To see comprehensively the Lifshitz effect on this $p$-wave superconductor model, it is helpful to extend our present calculations into the soliton background with the Lifshitz fixed point, which models the $p$-wave insulator/superconductor phase transition.  Furthermore, going beyond the probe limit, a rich phase structure was found for this $p$-wave model in the absence of the external magnetic field~\cite{Cai:2013aca,Li:2013rhw,Cai:2014ija}. Hence, in order to further understand the Lifshitz influence on the complete phase diagrams of this vector model, we will study the backreaction of the  MCV field on the Lifshitz black hole in the near future.

\acknowledgments  We would like to thank R.~G. Cai for his directive help. J.~W.~Lu is  deeply grateful to L.~Li for his helpful discussions and comments. This work is supported by the National Natural Science Foundation of China (Grants No. 11175077 and No. 11205078), the Ph.D Programs of the Ministry of China (Grant No. 20122136110002), and in part by a grant from State Key Laboratory of Theoretical Physics, Institute of Theoretical Physics, Chinese Academy of Science.
\appendix*
\section{Condensate of the complex vector field}
Because the charged vector field is dual to a vector operator in the boundary field theory, this MCV field can be regarded as an order parameter to model the $p$-wave superconductor phase transition. Here we consider the condensate of the MCV field in the absence of the external magnetic field.

Without loss of generality, we take the ansatz of the matter and gauge field sectors as
\begin{equation}
\rho_\nu dx^\nu=\rho_x(r) dx+\rho_y(r) dy, \ \ \ A_\nu dx^\nu=\phi(r) dt.
\end{equation}
Substituting the above ansatz into Eqs.~(\ref{EOMrho}) and (\ref{EOMphi}), we  get the following equations of motion:
\begin{eqnarray}
\phi''+\frac{3-z}{r}\phi'-\frac{2q^2}{r^4f}(\rho^2_x+\rho^2_y)\phi&=&0,\\
\rho_x''+\left(\frac{z+1}{r}+\frac{f'}{f}\right)\rho'_x+\left(\frac{q^2}{r^{2z+2}f^2}\phi^2-\frac{m^2}{r^2f}\right)\rho_x&=&0,\\
\rho_y''+\left(\frac{z+1}{r}+\frac{f'}{f}\right)\rho'_y+\left(\frac{q^2}{r^{2z+2}f^2}\phi^2-\frac{m^2}{r^2f}\right)\rho_y&=&0,
\end{eqnarray}
where the prime denotes the derivative with respect to $r$. If we choose  the $\rho_x$ component to condense, $\rho_y$ is imposed as vanishing, which leaves us with two coupled differential equations. To solve these equations, we first impose the boundary conditions. At the horizon, to ensure the finite form of $g^{\mu\nu}A_\mu A_\nu$, the gauge field should satisfy $\phi(r_+)=0$, while the vector field needs to be regular at the horizon. Near the boundary $r\rightarrow\infty$, the general falloffs of $\phi(r)$ and $\rho_x(r)$ read
\begin{eqnarray}
\phi(r)&=&\mu-\frac{\rho}{r^{2-z}}+\cdots,\\
\rho_x(r)&=&\frac{\rho_{x-}}{r^{\Delta_-}}+\frac{\rho_{x+}}{r^{\Delta_{+}}}+\cdots,
\end{eqnarray}
where $\Delta_\pm=\frac{1}{2}(z\pm\sqrt{z^2+4m^2})$. According to the gauge/gravity dual dictionary, $\mu$ and $\rho$ correspond to the chemical potential and the charge density in the dual field theory, while $\rho_{x-}$ and $\rho_{x+}$  correspond to the source and the vacuum expectation value of the boundary operator $J_x$, respectively. In order to meet the requirement that the symmetry of the system is spontaneously broken, we turn off the source term, i.e., $\rho_{x-}=0$.

There is an important symmetry in the system, which is of the form
\begin{equation}
r\rightarrow b r,\ \ \ \rho_{x+}\rightarrow b^{\Delta_++1}\rho_{x+},\ \ \ \rho\rightarrow b^2 \rho,\ \ T\rightarrow b^z T,
\end{equation}
with a constant $b$. By using this symmetry, we can fix the charge density of the system and then work in the canonical ensemble. As a special case of the phase transition induced by the external magnetic field, i.e., the case of $B=0$, we will take $\Delta_+=3/2$ and $q=1$ in the following calculations.

\begin{figure}
  \includegraphics[width=3.2 in]{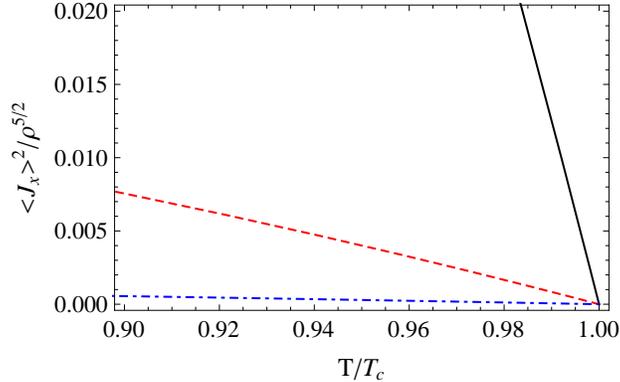}\\
  \caption{The condensate versus temperature with different $z$. The lines from top to bottom correspond to $z=1$~(black  solid),~$z=3/2$~(red  dashed),~and $z=9/5$~(blue  dot-dashed), respectively.}
  \label{conden}
\end{figure}

Near the critical temperature, the condensate $\langle J_x\rangle$ as a function of the temperature $T$ is plotted in Fig.~\ref{conden},
from which we find that there is a critical temperature $T_c$ for all cases ($z=1,~3/2,~9/5$), below which the vector field begins to condense. In particular, it turns out that the result in the case of $z=1$ is in accord with the one in Ref.~\cite{Cai:2013pda} by taking the square root of the value of  $\langle J_x\rangle^2/\rho^{5/2}$. The critical temperature for various $z$ is listed as follows:
\begin{equation}
z=1,\ \ T_c=0.102\rho^\frac{1}{2};\ \ \ \ z=\frac{3}{2},\ \ T_c=0.043\rho^\frac{3}{4};\ \ \ \ z=\frac{9}{5}, \ \ T_c=0.014\rho^\frac{9}{10}.
\end{equation}
From Fig.~\ref{conden} and the critical temperature, we see that the larger $z$ hinders the holographic superconductor phase transition, which is similar to the case with the excited Landau level. In addition, the condensate near $T_c$ for these three cases is fitted as
\begin{eqnarray}
z&=&1, \ \ \frac{\langle J_x\rangle^2}{\rho^\frac{5}{2}}=1.1870\left(1-\frac{T}{T_c}\right);\ \ \ \ z=\frac{3}{2}, \ \ \frac{\langle J_x\rangle^2}{\rho^\frac{5}{2}}=0.0807\left(1-\frac{T}{T_c}\right); \nonumber\\
z&=&\frac{9}{5},\ \  \frac{\langle J_x\rangle^2}{\rho^\frac{5}{2}}=0.0055\left(1-\frac{T}{T_c}\right).
\end{eqnarray}
Evidently, near the critical temperature, the linear dependence of $\langle J_x\rangle^2/\rho^{5/2}$ on $T$ indicates that the critical exponent $\frac{1}{2}$ is universal for all cases,  and thus the system undergoes a second-order transition. Besides, as $z$ increases, the coefficient of the condensate decreases, which agrees with the fact that $T_c$ decreases with $z$.

\end{document}